\documentclass[12pt]{article}
\usepackage{amssymb}
\newcommand{\eqd}{\stackrel{\rm def}{=}}

\newcommand{\normI}[1]{\mathop{\rm vrai\;sup}\limits_{{\bf q}\in
{\cal V}}\left|{#1}\right|}
\newcommand{\normII}[1]{\mathop{\rm vrai\;sup}\limits_{({\bf q},{\bf q}')
\in {\cal V}\!\times {\cal V}} \left|{#1}\right|}
\newcommand{\normIIqp}[1]{\mathop{\rm vrai\;sup}\limits_{({\bf q},{\bf p})
\in {\cal V}\!\times {\cal V}} \left|{#1}\right|}

\begin{document}

\thispagestyle{empty}

\author{M.I. Kalinin}
\title{On the completeness of describing an equilibrium canonical
ensemble using a pair distribution function}

\date{}
\maketitle

\begin{abstract}
It is shown that in equilibrium a canonical ensemble of particles
with two-particle interaction the Gibbs distribution function may
be expressed uniquely through a pair distribution function. It
means, that for given values of the particle number $N$, volume
$V$, and temperature $T$, the pair distribution function contains
as many information about the system as a full Gibbs distribution.
The latter is represented as a series expansion in the pair
distribution function. A recurrence relation system is
constructed, which allows all terms of this expansion to be
calculated successively.
\end{abstract}

\section{Introduction}

In classical statistical mechanics all properties of a closed
system of $N$ particles in a volume $V$ (canonical ensemble) are
described by a distribution function $D_N(X)$, where $X$ is a set
of system phase variables. For a system with an additive particle
interaction, multi-particle distribution functions are introduced
\cite{Balescu,NN} to calculate thermodynamical values of the
system. These multi-particle functions are considered to have less
information than the full Gibbs $N$-particle distribution
function. Thereby the smaller the order of the distribution
function is the less information it contains \cite{Balescu}. But
there does not exist a proof of this statement in the literature.

At the same time it is known that for an equilibrium system of $N$
non-interacting particles in the volume $V$ with the temperature
$T$ the GDF decomposes into a product of one-particle distribution
functions. This means that all information about the system is
contained in the one-particle distribution function.

In this article a problem on equivalence of describing an
equilibrium system by full Gibbs distribution function (GDF) and
pair (two-particle) distribution function (PDF) is considered. It
is used a non-normalized GDF and particle distribution functions
are defined trough it. The motive for that will be explained in
concluding section. As to a description of system the
non-normalized GDF contains all information about it.

We shall show that for the equilibrium system of $N$
pair-interacting particles in the absence of an external field,
when the energy $U_N$ equals
\begin{equation}
\label{energy} %
U_N({\bf q}_{1},\cdots ,{\bf q}_{N})= \sum_{1\le
i<j\le N}^{} u_{2}({\bf q}_i,{\bf q}_j),
\end{equation}                                
a Gibbs distribution may be expressed uniquely through a PDF. It
means the PDF contains all information on the system under
consideration.

\section{Formulation of the problem}

The equilibrium canonical ensemble is described by the Gibbs
distribution function\footnote{Here we consider only a
configuration part of the distribution function, because its
momentum part is factorized per a one-particle Maxwell function.}
\begin{equation}
\label{canon} %
D_N({\bf q}_1,\ldots,{\bf q}_N)= \exp\{-\beta
U_N({\bf q}_1,\ldots,{\bf q}_N)\},\quad {\bf q}_i\in {\cal V},
\quad i=1,\ldots,N,\quad\beta=1/kT.
\end{equation}                                          
where $k$ is the Boltzmann constant, ${\cal V}$ is a domain of
three-dimensional Euclidean space\footnote{Three-dimensionality of
space is not essential. It may be of any finite dimensionality.
Only finiteness of the Lebegue measure (volume) of the manifold
${\cal V}$ is important.}. Here and below in this paper we shall
suppose that the configuration variables ${\bf q}$ belong to the
domain ${\cal V}$.

For the system of particles with the energy (\ref{energy}) the
expression (\ref{canon}) may be rewritten as follows
\begin{equation}
\label{canon-v} %
D_N({\bf q}_1,\ldots,{\bf q}_N)= \prod_{1\le i<j\le N}v({\bf
q}_i,{\bf q}_j),
\end{equation}                                        
where we introduce the function
\begin{equation}
\label{v1v2} %
v({\bf q},{\bf q}')\eqd \exp\{-\beta u_2({\bf q},{\bf }')\}.
\end{equation}                                        
Due to reality of the potential $u_2$ this function is
nonnegative. And by virtue of symmetry with respect to its
arguments the function $v({\bf q},{\bf q}')$ is symmetrical too.
It follows from the expression (\ref{canon-v}) that statistical
properties of the system under consideration are described
completely by specifying a single function $v({\bf q},{\bf q}')$
of two configuration variables. Naturally, there appears an
assumption that, if this system may be described entirely with the
help of one function of two variables $v({\bf q},{\bf q}')$, then
it may also be described with the help of another function of two
variables connected with previous one, namely a pair distribution
function. We prove this assumption here.

Define $s$-particle distribution functions ${\cal G}_s$ starting
from the non-normalized GDF (\ref{canon-v})
\begin{equation}
\label{G_s} %
{\cal G}_s({\bf q}_1,\ldots,{\bf q}_s)=
\frac{N!}{(N-s)!V^N}\int D_N({\bf q}_1,\ldots,{\bf q}_N) d{\bf
q}_{s+1}\cdots d{\bf q}_N\quad s=1,\ldots N.
\end{equation}                                            
Here $V$ is a volume of the domain ${\cal V}$. The integration
will be carried out over the ${\cal V}$ and the domain of
integration will not be indicated explicitly.

Due to the properties of $v({\bf q},{\bf q}')$ all multi-particle
distribution functions are real, nonnegative, and symmetrical.
These functions are related to the ones $f_s$, introduced by
Balescu \cite{Balescu}, and to $F_s$ introduced by Bogoliubov
\cite{NN} as follows: $f_s=Q_N^{-1}V^N{\cal G}_s$ and $F_s
=(Q_NN!)^{-1}(N-s)!V^{N-s}{\cal G}_s$ respectively, where $Q_N$ is
a configuration integral
\begin{equation}
\label{conf_int} %
Q_N=\int\!d{\bf q}_1\cdots d{\bf q}_N D_N({\bf }_1,\ldots,{\bf
q}_N)=\frac{V^N(N-s)!}{N!}\int\!d{\bf q}_1\cdots d{\bf q}_s {\cal
G}_s({\bf }_1,\ldots,{\bf q}_s).
\end{equation}                                       

Write down explicitly an expression for the PDF
\begin{equation}
\label{G_2} %
{\cal G}_2({\bf q}_1,{\bf q}_2)=
\frac{N(N-1)}{V^N}\int\prod_{1\le i<j\le N}^{} v({\bf q}_i,{\bf
q}_j)d{\bf q}_3\cdots d{\bf q}_N.
\end{equation}                                            
Let us introduce a nonlinear operator $\Phi$ as follows
\begin{equation}
\label{Phi} %
\Phi(v)\eqd\frac{1}{V^{N-2}}\int\prod_{1\le i<j\le
N}^{} v({\bf q}_i,{\bf q}_j)d{\bf q}_3\cdots d{\bf  q}_N.
\end{equation}                                            
Then the relation (\ref{G_2}) may be considered as a nonlinear
operator equation relative to $v$
\begin{equation}
\label{F(G_2,v)} %
\frac{V^2}{N(N-1)}{\cal G}_2-\Phi(v)\eqd{\cal F}({v,\cal G}_2)=0.
\end{equation}                                   
If this equation has a solution
\begin{equation}
\label{chi(G_2)} %
v=\chi({\cal G}_2),
\end{equation}                                    
then its substitution into (\ref{canon-v}) permits one to express
the non-normalized Gibbs distribution $D_N({\bf q}_1,\ldots,{\bf
q}_N)$ through the two-particle function ${\cal G}_2({\bf q},{\bf
q}')$. Accordingly all $s$-particle functions may also be
expressed via ${\cal G}_2({\bf q},{\bf q}')$.

For calculating the function $v$ from equation (\ref{F(G_2,v)}) it
is necessary to set an additional condition
\begin{equation}
\label{G_2/Phi^0} %
{\cal G}_2^{(0)}=\frac{N(N-1)}{V^2}\Phi(v^{(0)}),
\end{equation}                                       
where ${\cal G}_2^{(0)}$ and $v^{(0)}$ are known functions. It is
easy to find such a condition assuming particularly
\begin{equation}
\label{v^0G^0} %
v^{(0)}({\bf q},{\bf q}')=1,\qquad {\cal G}_2^{(0)}({\bf q},{\bf
q}')=\frac{N(N-1)}{V^2},
\end{equation}                                       
which corresponds to the choice of noninteracting particle system
as a reference one\footnote{Evidently, instead of unity for
$v^{(0)}({\bf q},{\bf q}')$ we may take an arbitrary constant $C$.
In this case an expression for ${\cal G}_2^{(0)}({\bf q},{\bf
q}')$ is calculated easily too.}.

\section{Proof of the existence of a solution}

In functional analysis there exist a number of theorems about
implicit functions for operators of different smoothness degree
(e.g. \cite{Vainberg,Kant}). We use a theorem for analytical
operators in Banach spaces in the form presented in the book
\cite[Theorem 22.2]{Vainberg}.

{\bf Theorem.} {\it Let ${\cal F}(v,{\cal G}_2)$ be an analytical
operator in $D_r(v^{(0)},E_1)\dot{+}D_{\rho}({\cal G}_2^{(0)},E)$
with values in $E_2$. Let the operator $B\eqd -\partial{\cal
F}(v^{(0)},{\cal G}_2^{(0)})/\partial v$ have a bounded inverse
operator. Then there exist such positive numbers $\rho_1$ and
$r_1$ that in the solid sphere $D_{r_1}(v^{(0)},E_1)$ there exists
a unique solution (\ref{chi(G_2)}) to equation (\ref{F(G_2,v)}).
This solution is defined in the solid sphere $D_{\rho_1}({\cal
G}_2^{(0)},E)$, being analytical in it and satisfying the
condition $v^{(0)}=\chi({\cal G}_2^{(0)})$.}

\smallskip\noindent

Here $D_r(x_0,{\cal E})$ is a solid sphere of the radius $r$ in
the vicinity of an element $x_0$ of Banach space ${\cal E}$, the
sign "$\dot{+}$" denotes an algebraic sum of manifolds, and
$\partial{\cal F}/\partial v$ is a Frechet derivative
\cite{Kant,Vulih} of a nonlinear operator ${\cal F}$.

To investigate the problem of existence and uniqueness of the
solution of equation (\ref{F(G_2,v)}) it is necessary to verify
fulfillment of the conditions of the theorem for the system under
consideration.

At first define spaces $E_1$ and $E$, in which the operator ${\cal
F}$ is set, and space $E_2$ containing a range of values of this
operator. We shall use a class of functions $v({\bf q},{\bf q}')$
bounded almost everywhere in ${\cal V}$. The space of such
functions becomes a complete normalized (Banach) space of
essentially bounded functions $L_{\infty}$ if we introduce the
following norm \cite{Kant, Vulih}
\begin{equation}
\label{norm} %
\|v\|=\mathop{\rm vrai\,sup}\limits_{({\bf q},{\bf q}') \in {\cal
V}\!\times {\cal V}}|v({\bf q},{\bf q}')|,
\end{equation}                                          
where "$\mathop{\rm vrai\,sup}$" denotes an essential exact
supremum of the function on the indicated set, and ${\cal
V}\!{\times}{\cal V}$ is a direct product of the manifold ${\cal
V}$ by itself, that is a set of ordered pairs $({\bf q},{\bf
q}')$. Moreover $v({\bf q},{\bf q}')$ is a symmetrical function of
its arguments. Therefore $v({\bf q},{\bf q}')\in
L^{(s)}_{\infty}({\cal V}\!{\times}{\cal V})$. The index $(s)$
means that we use a space of symmetrical functions. Such
properties of the function $v({\bf q},{\bf q}')$ defined by
(\ref{v1v2}) are provided by any interaction potential bounded
below almost everywhere in the domain ${\cal V}$. All potentials
conventionally used in statistical mechanics satisfy this
condition.

Using the properties of real functions and their integrals
\cite{Vulih}, one can show that for any functions $v$ from
$L^{(s)}_{\infty}({\cal V}\!{\times}{\cal V})$ both GDF and all
multi-particle ones are also essentially bounded functions being
defined on appropriate sets. Thus the space of functions $v$ is
$E_1= L^{(s)}_{\infty}({\cal V}\!{\times}{\cal V})$, the space of
functions ${\cal G}_2$ is $E=L^{(s)}_{\infty}({\cal V}
\!{\times}{\cal V})$, and the space containing the range of values
of the operator ${\cal F}$ is $E_2=L^{(s)}_{\infty}({\cal V}
\!{\times}{\cal V})$. The set of pairs $(v,{\cal G}_2)$ of
nonnegative functions in the direct product $E_1\times E$ is a
range of definition\footnote{In general, the requirement on
nonnegativity of the functions follows from physical reasons. From
the mathematical point of view the operator ${\cal F}(v,{\cal
G}_2)$ (\ref{F(G_2,v)}) is defined on a whole space $E_1\times
E$.} of the operator ${\cal F}$, with its range of values
belonging to $E_2$. Everywhere in the range of definition the
operator ${\cal F}(v,{\cal G}_2)$ is analytical.

Calculate an operator $B$ defined in the theorem formulation for
chosen values of the functions $v^{(0)}$ and ${\cal G}_2^{(0)}$
(\ref{v^0G^0}). It is a linear operator acting from space $E_1$ to
space $E_2$. Using the first equality in (\ref{F(G_2,v)}), we
obtain the expression
\begin{equation}
\label{B-def} %
B=\frac{d\Phi(v^{(0)})}{dv}.
\end{equation}                               

Substitute into it the expression (\ref{Phi}) for the operator
$\Phi(v)$ and use the condition $v^{(0)}=1$. We obtain the
operator $B$ as follows
\begin{equation}
\label{B} %
(Bh)({\bf q},{\bf q}')=h({\bf q},{\bf q}')+
(N-2)\left[\bar{h}({\bf q})+\bar{h}({\bf  q}')\right]+
\frac{(N-2)(N-3)}{2}\bar{\bar{h}},
\end{equation}                                      
where we introduce designations
\begin{equation}
\label{h1h2} %
\bar{h}({\bf q})=\frac{1}{V}\int h({\bf q},{\bf q}')d{\bf q}',
\quad \bar{\bar{h}}=\frac{1}{V^2}\int h({\bf q},{\bf q}')d{\bf
q}d{\bf q}'.
\end{equation}                                            
In these expressions $h\in E_1$, $Bh\in E_2$. The range of
definition of $B$ is a whole space $E_1$: $D(B)=E_1$ and its range
of values is a space $E_2$: $R(B)=E_2$. The operator $B$ is
bounded. Its norm defined by (see eg. \cite{Kant, Vulih})
\begin{equation}
\label{||B||} %
\|B\|=\mathop{\rm sup}\limits_{\{h\in E_1;\|h\|=1\}}\|Bh\|,
\end{equation}                                      
is evaluated easily with the help of (\ref{B}). Using properties
of norms we obtain the following inequality
\begin{equation}
\label{||Bh||<I} %
\|Bh\|\le\|h({\bf q},{\bf q}')\|+(N-2)\|\bar{h}({\bf q})+
\bar{h}({\bf q}')\|+\frac{(N-2)(N-3)}{2}\|\bar{\bar{h}}\|.
\end{equation}                                      

Evaluate the norms in right hand side of (\ref{||Bh||<I}). The
value $\bar{\bar{h}}$ is a constant. For its norm we have an
estimate
$$
\|\bar{\bar{h}}\|=\left|\frac{1}{V^2}\int h({\bf q},{\bf q}')
d{\bf q}d{\bf q}'\right|\leq\frac{1}{V^2}\int \normII{h({\bf
q},{\bf q}')} d{\bf q}d{\bf q}'=
$$
\begin{equation}
\label{hhh} %
=\frac{1}{V^2}\int \|h\|d{\bf q}d{\bf q}'= \|h\|.
\end{equation}                                      
For the norm of a sum of the functions $\bar{h}({\bf q})$ and
$\bar{h}({\bf q}')$ we obtain the following estimate
$$
\|\bar{h}({\bf q})+\bar{h}({\bf q}')\|=\normII{\bar{h}({\bf q})
+\bar{h}({\bf q}')} \leq 2\normI{\bar{h}({\bf q})}=
$$
$$
=\frac{2}{V}\normI{\int h({\bf q},{\bf p}) d{\bf
p}}\le\frac{2}{V}\int \normI{h({\bf q},{\bf p})} d{\bf p}\leq
$$
\begin{equation}
\label{hh} %
\leq\frac{2}{V}\int \normIIqp{h({\bf q},{\bf p})}d{\bf p} =2\|h\|.
\end{equation}                                      

Substituting (\ref{hhh}) and (\ref{hh}) in the inequality
(\ref{||Bh||<I}), we obtain the following estimate for the norm of
the function $(Bh)({\bf q},{\bf q}')$
\begin{equation}
\label{||Bh||<II} %
\|Bh\|\leq\frac{N(N-1)}{2}\|h\|.
\end{equation}                                      
Taking into account the definition (\ref{||B||}), we obtain an
estimate of the norm of the operator $B$
\begin{equation}
\label{||B||<} %
\|B\|\leq\frac{N(N-1)}{2}.
\end{equation}                                      

One can show that actually in (\ref{||B||<}) the sign of equality
is valid. Indeed because there is at least one function $h({\bf
q},{\bf q}')\in E_1$ with the norm equal to unity for which
$\|Bh\|$ equals the value in the right hand side of
(\ref{||Bh||<II}), namely $h({\bf q},{\bf q}')\equiv 1$, then the
inequality (\ref{||B||<}) reduces to the equality
\begin{equation}
\label{||B||=} %
\|B\|=\frac{N(N-1)}{2}.
\end{equation}                                      

Now define a kernel (a space of zeros) of the operator $B$, i.e.
find a solution to the homogeneous equation $Bh=0$. Using
(\ref{B}), we obtain the equation
\begin{equation}
\label{Bh=0} %
h({\bf q},{\bf q}')+(N-2)[\bar{h}({\bf q})+\bar{h}({\bf q}')]+
\frac{1}{2}(N-2)(N-3)\bar{\bar{h}}=0.
\end{equation}                                      
Integration of this equation over ${\bf q}'$ and then over ${\bf
q}$ allows one to obtain its solution. It turns out to be trivial
$h({\bf q},{\bf q}')=0$.

This means that the corresponding inhomogeneous equation $Bh=g$ is
uniquely solvable. Its solution has the form
\begin{equation}
\label{h=B^{-1}g} %
h({\bf q},{\bf q}')\eqd(B^{-1}g)({\bf q},{\bf q}')= g({\bf q},{\bf
q}')-\frac{N-2}{N-1}[\bar{g}({\bf q})+ \bar{g}({\bf
q}')]+\frac{N-2}{N}\bar{\bar{g}},
\end{equation}                                      
where $\bar{g}({\bf q})$ and $\bar{\bar{g}}$ are determined by the
formulae analogous to (\ref{h1h2}) where $h$ is replaced by $g$.
It is easy to see that the operator $B^{-1}$ is specified for all
$g\in E_2$, that is indeed $R(B)=E_2$. It is easy to verify that
the conditions $B^{-1}B=BB^{-1}=I$ are valid. In other words, the
operator $B$ is both left and right inverse. Therefore it is the
inverse operator to $B$ \cite{Trenogin}.

The operator $B^{-1}$ is bounded. For its norm an upper bound is
easily evaluated just as for the operator $B$. At once write down
this estimate
\begin{equation}
\label{||B^{-1}||<} %
\|B^{-1}\|\le \frac{3N^2-6N+2}{N(N-1)}\:.
\end{equation}                                      

Thus all conditions of the theorem are fulfilled. Hence there is a
unique solution (\ref{chi(G_2)}) expressing unambiguously the
function $v({\bf q},{\bf q}')$ in terms of PDF ${\cal G}_2({\bf
q},{\bf q}')$. This solution is determined in the solid sphere
$D_{\rho_1}({\cal G}_2^{(0)},E)$, analytical in it, and satisfies
the conditions (\ref{G_2/Phi^0}), (\ref{v^0G^0}). We shall not
produce here a proof of this theorem. It is presented in many
books in functional analysis. Indicate merely the procedure of
constructing a solution to equation (\ref{F(G_2,v)}) and
estimating the convergence domain size.

Introduce functions $g({\bf q},{\bf q}')$ and $h({\bf q},{\bf
q}')$ by the relations
\begin{equation}
\label{g-h} %
{\cal G}_2=\frac{N(N-1)}{V^2}(1+g), \qquad  v=1+h.
\end{equation}                                             
Here and below $h({\bf q},{\bf q}')$ is a Mayer function. Equation
(\ref{F(G_2,v)}) with the operator $\Phi(v)$ (\ref{Phi}) may be
rewritten as
\begin{equation}
\label{h=Psi(g,h)} %
h=B^{-1}g+B^{-1}\sum_{s=2}^{\cal N}B_s(h),
\end{equation}                                      
where ${\cal N}=N(N-1)/2$ and $B_s(h)$ is a homogeneous power
operator of the order $s$ with respect to $h$ equal to
\begin{equation}
\label{B_sh^s} %
B_s(h)=-\frac{1}{V^{N-2}}\int d{\bf q}_3\cdots d{\bf  q}_N
\sum_{1\leq K_1<\cdots<K_s\leq {\cal N}}h(X_{K_1})\cdots
h(X_{K_s}).
\end{equation}                                           
Here we introduce a generalized index $K$ varying from 1 to $\cal
N$ enumerating all possible different index pairs $(i\,j)$,
satisfying the condition $i<j$ which can be constructed from the
numbers $1,2,\ldots,N$. Such one-to-one correspondence
$K\leftrightarrow (i\,j)$ is easily found. $X_K$ designates the
corresponding pair of coordinates $({\bf q}_i,{\bf q}_j)$. One can
consider that the power operator $B_s(h)$ is generated by the
$s$-linear operator $R_s(h_1,\ldots,h_s)$ \cite{Vainberg}
\begin{equation}
\label{R_s(h^s)} %
R_s(h_1,\ldots,h_s)=-\frac{1}{V^{-2}N}\int d{\bf
q}_3\cdots d{\bf  q}_N \sum_{1\leq K_1<\cdots<K_s\leq {\cal N}}
h_1(X_{K_1})\cdots h_s(X_{K_s}).
\end{equation}                                        
Using this operator, we rewrite equation (\ref{h=Psi(g,h)}) as
follows
\begin{equation}
\label{h==Psi(g,h)} %
h=B^{-1}g+B^{-1}\sum_{s=2}^{\cal N}R_s(h,\ldots,h).
\end{equation}                                      
Its solution is found as a power series in $g$
\begin{equation}
\label{h(g)} %
h=\sum_{k=1}^{\infty}h_k(g),
\end{equation}                                            
where $h_k(g)$ is an homogeneous power operator of the order $k$
acting from space $E_2$ to $E_1$. This means that each operator
$h_k(g)$ is a function of two configuration variables.

Substituting the expression (\ref{h(g)}) into equation
(\ref{h==Psi(g,h)}) and producing some rearrangement of terms in
the right-hand side, we derive the following relationship
\begin{equation}
\label{ext1} %
\sum_{k=1}^{\infty}h_k(g)=B^{-1}g+B^{-1}\sum_{k=2}^{\infty}
\sum_{s=2}^{k}\theta_{s{\cal N}}\sum_{k_1+\cdots+k_s=k}
R_s(h_{k_1}(g),\ldots,h_{k_s}(g)),
\end{equation}                                      
where
$$\theta_{s{\cal N}}=\left\{\begin{array}{ll}
1, & s\leq{\cal N},\\
0, & s>{\cal N}.
\end{array}\right.
$$
In accordance with the theorem on uniqueness of analytical
operators \cite{Hille}, equating terms of identical orders on $g$,
we obtain a recurrent system for determining the operators
$h_k(g)$
$$
h_1(g)=B^{-1}g,
$$
\begin{equation}
\label{h_kg^k} %
h_k(g)=B^{-1}\sum_{s=2}^{k}\theta_{s{\cal N}}
\sum_{k_1+\cdots+k_s=k} R_s(h_{k_1},\ldots,h_{k_s}),\quad
k=2,3,\ldots.
\end{equation}                                      
All expansion terms of (\ref{h(g)}) are calculated successively
from this system. And thus a formal solution $h(g)$ is derived in
the form of the above series (\ref{h(g)}). It is also necessary to
prove a convergence of this series. It is established, just as in
\cite{Vainberg}, by Goursat method \cite{Goursat} of majorizing
functions.

Consider a subsidiary equation
\begin{equation}
\label{major} %
\xi=\gamma\left\{\eta+\sum_{s=2}^{\cal N} \frac{{\cal
N}!}{s!\,({\cal N}-s)!}\;\xi^s\right\},
\end{equation}                                      
in which we designate $\xi=\|h\|$, $\eta=\|g\|$ and
$\gamma=\|B^{-1}\|$. The right-hand side of this equation
majorizes the norm of the right-hand side of (\ref{h==Psi(g,h)}).
A solution to equation (\ref{major}) can be found in the form of
expansion in series with respect to $\eta$:
$\xi=\sum_{k=1}^{\infty}\xi_k \eta^k$. For coefficients $\xi_k$ it
is easy to derive a recurrent system analogous to the structure of
(\ref{h_kg^k}). Just as in a classical case \cite{Goursat} it is
easy to show that $\|h_k(g)\|\le \xi_k \eta^k$. Therefore if the
series for $\xi$ converges, then the series (\ref{h(g)})
absolutely converges.

On the other hand, equation (\ref{major}) can be rewritten as
\begin{equation}
\label{major-2} %
\gamma\eta=\xi-\gamma[(1+\xi)^{\cal N}-1-{\cal N}\xi].
\end{equation}                                      
We cannot write out a solution to this equation explicitly by
quadratures, but it is easy to show graphically that the solution
$\xi(\eta)$ satisfying the condition $\xi(0)=0$ exists in the
region $\eta<\eta_0$. This solution is unique, analytical and
monotonically increasing. The value $\xi_0$ determines the maximum
point of the function in the right-hand side of equation
(\ref{major-2}) and equals $\xi_0=(1+1/\gamma{\cal
N})^{\frac{1}{{\cal N}-1}}$. The value $\eta_0$ is defined by this
function maximum and equals
$$
\eta_0=({\cal N}-1)\left(1+\frac{1}{\gamma{\cal N}}\right)
\left[\left(1+\frac{1}{\gamma{\cal N}}\right)^{\frac{1}{{\cal N}
-1}}-1\right]-\frac{1}{\gamma{\cal N}}.
$$
Thus the series (\ref{h(g)}) converges if $\|g\|<\eta_0$. Thereby,
as a radius $\rho_1$ of the solid sphere from the theorem, any
number smaller than $\eta_0$ may be chosen. By virtue of a
monotonic increase of the function $\xi(\eta)$, the radius  $r_1$
may be taken equal to $r_1=\xi(\rho_1)$. So the solution
(\ref{chi(G_2)}) to the problem (\ref{F(G_2,v)}) exists, it is
unique and may be presented in the form of the expansion
(\ref{h(g)}) into an absolutely convergent series with respect to
$g=({\cal G}_2-{\cal G}_2^{(0)})/{\cal G}_2^{(0)}$ at least in the
region $\|g\|<\eta_0$.

Since the Gibbs distribution (\ref{canon-v}) is a polynomial of
the order ${\cal N}$ relative to the function $h$, then it is
analytical function of $g$ and may be expanded in a power series
with respect to $g$
\begin{equation}
\label{D_N(g)} %
D_N=\sum_{k=0}^{\infty}d_k(g),
\end{equation}                                       
where $d_k(g)$ is a homogeneous power operator of the order $k$
acting from a space $E_2$ into a space  $E_N=
L^{(s)}_{\infty}(\underbrace{{\cal V}\!{\times}\cdots{\times}
{\cal V}}_{\mbox{\tiny{$N$  co-factors}}})$, that is, each
operator $d_k(g)$ is a function of $N$ configuration variables.
Due to determining the PDF (\ref{G_2}) and the theorem on
uniqueness of analytical operators \cite{Hille} these functions
satisfy the conditions
$$
d_0=1,
$$
\begin{equation}
\label{b1_int} %
\frac{1}{V^{N-2}}\int\!d{\bf q}_3\cdots d{\bf q}_N d_1({\bf
q}_1,\ldots,{\bf q}_N)=g(q_1),
\end{equation}                                       
$$
\int\!d{\bf q}_1\cdots d{\bf q}_N d_k({\bf q}_1,\ldots,{\bf
q}_N)=0,\qquad k=2,3,\ldots .
$$

Substituting the expansion (\ref{h(g)}) into into (\ref{canon-v})
and producing some rearrangement of terms, we can derive uniquely
all terms $d_k(g)$ of the expansion(\ref{D_N(g)}) in the form
\begin{equation}
\label{d_k} %
d_k(g)=\sum_{s=1}^{k}\theta_{s{\cal N}}\sum_{k_1+
\cdots+<k_s=k}\sum_{1\leq K_1<\cdots<K_s\leq {\cal N}}
h_{k_1}(X_{K_1})\cdots h_{k_s}(X_{K_s}),\quad k=1,2,\ldots,
\end{equation}                                      
where $h_i(X_{K_i})$ are expressed in terms of $g$ by formulae
(\ref{h_kg^k}). The expressions obtained determine an expansion of
the GDF in  series with respect to PDF.

\section{Concluding remarks}

Using the theorem of existence and uniqueness of the solution to
equation (\ref{F(G_2,v)}), we reveal that Mayer function $h({\bf
q},{\bf q'})$ may be expressed uniquely through the PDF ${\cal
G}_2({\bf q},{\bf q'})$ (\ref{G_2}). It permits us to present the
non-normalized Gibbs distribution as a nonlinear operator function
of PDF and write its expansion to power series. The procedure of
successive calculation of the all terms of this series is
formulated.

So knowing a pair distribution function, we can express the Gibbs
distribution uniquely in terms of this function. In turn PDF is
expressed uniquely via GDF (see (\ref{G_s})). This means that
descriptions of equilibrium canonical ensemble by both GDF and PDF
are equivalent, at least for $h$ belonging to the solid sphere
$D_{r_1}(0,E_1)$. Hence PDF contains the same quantity of
information as the full GDF, as well as all $s$-particle
distribution functions for $s>2$ which are also expressed uniquely
through PDF.

The one-particle distribution function ${\cal G}_1({\bf q})$ does
not seem to contain all information about the system with
two-particle interaction. If one tries to realize the above
programme, then one fails to satisfy all conditions of the theorem
on existence and uniqueness of the solution. The kernel (a space
of zeros) of the corresponding operator $B$ consists of arbitrary
functions $h({\bf q},{\bf q}')\in L_{\infty}^{(s)} ({\cal
V}\times{\cal V})$ limited only by one condition $\bar{h}({\bf
q})\equiv 0$. Therefore it is possible for such operator $B$ to
have no inverse operator. And indeed it has not.

Note here that for the normalized GDF an analogous programme
cannot be realized too. In this case the corresponding operator
$B$ has also a nontrivial kernel, namely a set of all functions
constant on ${\cal V}$. Mathematically, it is clear from the form
of the normalized GDF in which the function $v$ may be multiplied
by an arbitrary constant without changing the distribution
function. Physically, it corresponds to the fact that the
interacting potential is defined with an accuracy to an arbitrary
constant.


\begin{thebibliography}{99}
\bibitem{Balescu}
R. Balescu. Equilibrium and nonequilibrium statistical mechanics.
John Wiley \& Sons Inc. N.Y., 1975.
\bibitem{NN}
N.N. Bogoliubov. Problems of dynamical theory in statistical
physics. In Selected proceedings, Vol. 2, pp. 99-196, 1970 (in
Russian).
\bibitem{Vainberg}
M.M. Vainberg, and V.A. Trenogin. A Theory of branching of
nonlinear equation solutions. M., Nauka, 1969, (in Russian).
\bibitem{Kant}
L.V. Kantorovich, and G.P. Akilov. Functional analysis. M., Nauka,
1977 (in Russian).
\bibitem{Vulih}
B.Z. Vulikh. A brief course in the theory of functions of real
variables (An introduction to the theory of integral). M., Mir,
1976.
\bibitem{Trenogin}
V.A. Trenogin. Functional analysis. M., Nauka, 1980 (in Russian).
\bibitem{Hille}
E. Hille, and R.S. Phillips. Functional analysis and semi-groups.
Providence, 1957.
\bibitem{Goursat}
E. Goursat. Course of mathematical analysis, Vol. 1. M.-L., ONTI,
1936 (in Russian); (E. Goursat. Cours d'analyse math\'ematique,
Vol. 1. Gauthier - Villars, Paris, 1923).

\end{thebibliography}
\end{document}